\documentclass[twocolumn,secnumarabic,amssymb, nobibnotes, aps, prd,longbibliography]{revtex4-1}
\pdfoutput=1

\usepackage{graphicx}% Include figure files
\usepackage{caption}
\usepackage{gensymb}
\usepackage{braket}
\usepackage{subcaption}
\usepackage{amsmath}
\usepackage{textcomp}
\usepackage{color}
\usepackage{amsmath}
\usepackage{braket}
\usepackage{dcolumn}
\usepackage{graphicx}
\usepackage{lipsum}
\usepackage{wrapfig}
\usepackage{gensymb}
\usepackage{textcomp}
\usepackage{graphicx}% Include figure files
\usepackage{caption}
\usepackage{color}
\usepackage{dcolumn}
\usepackage{wrapfig}
\begin{document}
\title{Nanoscale electrical conductivity imaging using a nitrogen-vacancy center in diamond}%

\author{Amila Ariyaratne, Dolev Bluvstein, Bryan A. Myers \& Ania C. Bleszynski Jayich}%
\email[email: ]{ania@physics.ucsb.edu}
\affiliation{Department of Physics and Astronomy\\ 
University of California Santa Barbara,
Santa Barbara, CA 93106}

\begin{abstract}
The electrical conductivity of a material can feature subtle, nontrivial, and spatially-varying signatures with critical insight into the material's underlying physics. Here we demonstrate a conductivity imaging technique based on the atom-sized nitrogen-vacancy (NV) defect in diamond that offers local, quantitative, and noninvasive conductivity imaging with nanoscale spatial resolution. We monitor the spin relaxation rate of a single NV center in a scanning probe geometry to quantitatively image the magnetic fluctuations produced by thermal electron motion in nanopatterned metallic conductors. We achieve 40-nm scale spatial resolution of the conductivity and realize a 25-fold increase in imaging speed by implementing spin-to-charge conversion readout of a shallow NV center. NV-based conductivity imaging can probe condensed-matter systems in a new regime, and as a model example, we project readily achievable imaging of nanoscale phase separation in complex oxides.
\end{abstract}

\maketitle

The motion of electrons in the solid-state provides important insight into a material's multiple interacting degrees of freedom, and understanding the complexity of these interactions is at the heart of condensed matter physics. Measurements of a material's conductivity and its signatures as a function of temperature or magnetic field, for instance, often provide the best evidence for the microscopic mechanisms at play. In recent years, significant interest has turned to condensed matter phenomena in which the electrical conductivity exhibits nontrivial spatial variations. Examples include topological insulators \cite{Moore:2010ig,Langbehn2017}, which host conducting surfaces and an insulating interior, and Mott insulators that at critical temperatures and magnetic fields exhibit nanoscale phase separation, where pocket-like metallic regions form in an insulating matrix \cite{Mattoni2016,Imada1998}. As another example, anomalous domain wall conductivity has been observed in various complex oxides, such as multiferroics \cite{Seidel2010} and  iridates \cite{Ma2015,Ueda2014a}, with predictions of exotic Weyl semimetal behavior at these domain walls \cite{Yamaji:2014fa}. These phenomena are challenging to probe with standard transport measurements that average over a macroscopic area of the sample, and detecting these spatial signatures is the goal of many advanced techniques.

Many spatially-resolved probes of local electron dynamics exist, including optical conductivity probes and scanning probe microscopy (SPM)-based probes. Optical probes are limited in their spatial resolution by the diffraction limit (with specialized near-field probes approaching resolutions of tens of nanometers \cite{Atkin:2012cm,Fei:2012bi}), probe conductivity at optical frequencies only, and are generally surface sensitive. SPM-style probes, such as conducting atomic force microscopy (AFM) \cite{Murrell1993} and microwave impedance microscopy (MIM) \cite{Lai2009}, offer high spatial resolution but are limited in other respects. A prominent drawback is that the signal they produce is convolved with the geometry of the probe and its interface with the sample under study, thus making quantitative measurements challenging. Importantly, all of these local conductivity probes measure the response of a system to some driving perturbation, which could induce a nonlinear response or void the subtle effects one hoped to study \cite{Altshuler2009}.

\begin{figure*}
\includegraphics[width=151.2mm]{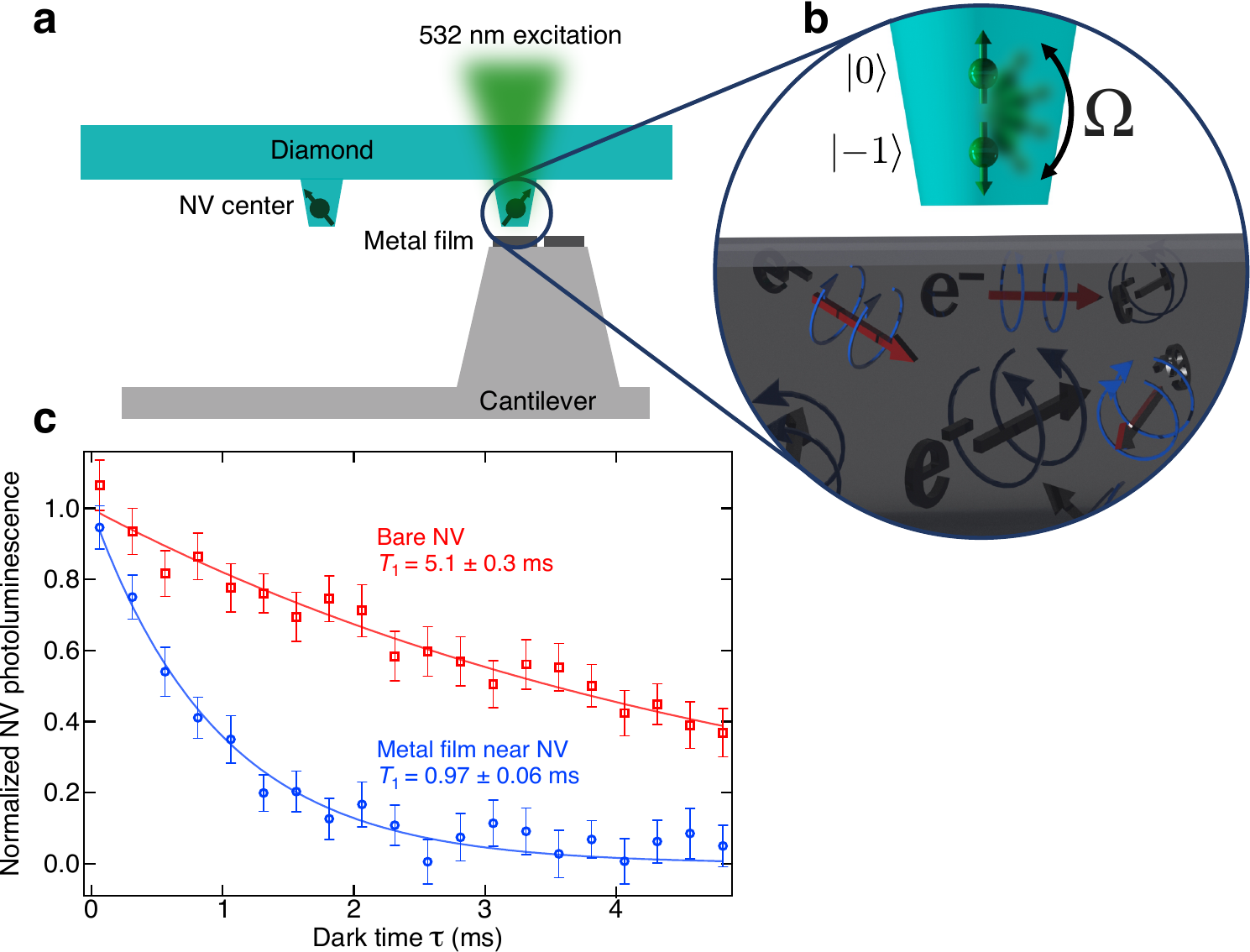}
\captionsetup{justification=raggedright, singlelinecheck=false}
\caption{Conductivity imaging using the nitrogen-vacancy (NV) center in diamond. \textbf{a} A metallic sample is deposited onto the flat tip of an in-house-fabricated silicon scanning probe. The flat plateau region is several micrometers in diameter. This `sample on tip' is scanned over a diamond pillar containing a single NV center. 532 nm excitation is used for optical control and readout of the NV spin triplet. \textbf{b} Illustration demonstrating NV spin relaxation in the presence of a conductor. The stochastic, thermal motion of free electrons produces magnetic fluctuations that increase the spin transition rate $\Omega$ between $|m_s=0\rangle$ and $|m_s=-1\rangle$, detectable as a reduction of the NV center's spin relaxation time $T_1$.
\textbf{c} Measurement of the $T_1$ of an NV center far from any conductor (red squares) and positioned 100 nm above the surface of an 85-nm thick Ag film (blue circles). The specific measurement sequence is discussed in the main text and yields an exponential photoluminescence decay $\exp{\left(-\tau/T_1\right)}$ with $T_1 = 1/3\Omega$. The presence of the Ag film reduces the NV $T_1$ 5-fold. Error bars correspond to measured standard error.}
\label{Setup}
\end{figure*}

On the other hand, the nitrogen-vacancy (NV) center in diamond presents a quantitative, noninvasive, and nanoscale sensor capable of measuring electrical conductivity via directly sensing the magnetic fields produced by thermal electron motion in the material. An atom-like quantum sensor, the NV center's signal is analytically related to the sample conductivity through fundamental constants. And because the NV monitors the fluctuating magnetic fields produced by electrons in thermal equilibrium, the NV sensing mechanism involves neither a driving perturbation nor sample contact, which also allows for subsurface sensing. Third, due to its atomic-size, the NV center affords very high spatial resolution \cite{Taylor2008,Rugar2014}, with recent magnetic imaging reaching sub-10 nm resolution \cite{Pelliccione2016,Haberle2015}. Lastly, the versatility of the NV center is an attractive feature; it is capable of operating over a wide range of temperatures \cite{Bar-Gill2013,Toyli2012} and in a wide range of imaging modalities, offering the possibility of combining conductivity imaging with simultaneous magnetic, electric, and thermal imaging. Recent work by Kolkowitz \textit{et al.} used the NV center to sense spectral and thermal signatures of electron behavior in metallic films, elucidating transitions from diffusive to ballistic transport regimes with changes in temperature \cite{Kolkowitz2015}.

	In this work we integrate an NV sensor with a scanning probe microscope to spatially image local electron conductivity in nanopatterned metal films with 40-nm scale spatial resolution. By monitoring the relaxation rate of the NV center spin state while the NV is scanned in nanometer proximity to a metal, we quantitatively measure the local conductivity of several metals. By using spin-to-charge conversion readout techniques on a shallow NV center, we demonstrate a 25-fold reduction in imaging time. With the sensitivity and spatial resolution demonstrated here, we project NV-based conductivity imaging of nanoscale phase separation in complex oxides, a model example of a spatially inhomogeneous phenomenon in a condensed matter system.  
\subsection*{Results}
\textbf{Relaxation model and experimental setup.} The NV center in diamond is a point defect comprising a substitutional nitrogen atom and a nearby vacancy in the carbon lattice. The two unpaired electrons of the defect center form a ground state spin triplet that features long energy relaxation times ($T_1  \sim \text{ms}$) at room temperature.  The conductivity imaging experiments described here utilize the sensitivity of the spin $T_1$ to fluctuating magnetic fields produced by electrons moving in a nearby conductor. 
   
   The Hamiltonian of the NV ground state spin triplet in the presence of a magnetic field $B$ is given by

\begin{flalign}
 H = \Delta S_{z}^{2} + \gamma B_{z}S_{z} + \frac{1}{2}\gamma(B_{x} - iB_{y})S_{+} + \frac{1}{2}\gamma(B_{x} + iB_{y})S_{-}
\label{hamiltonian}
\end{flalign}

where $\Delta=2.87$ GHz is the zero field splitting between the $|{m_{s}=0}\rangle$ and $|{m_{s}=\pm 1}\rangle$ triplet states, $\gamma$ is the electron gyromagnetic ratio, $S_{i}$ are the spin operators and $S_{\pm}$ are the spin raising and lowering operators $S_x \pm i S_y$, and the $\hat{z}$ direction is chosen to point along the NV axis \cite{Doherty2012}. Incoherently oscillating magnetic fields that are at the frequency of the $|{m_{s}=0 \rightarrow \pm 1}\rangle$ transition and perpendicular to the NV axis induce transitions between the two states, thus speeding up the NV center's relaxation rate \cite{Schafer-Nolte2014}. This sensitivity to fluctuating magnetic fields can then be used to detect stochastic electron motion in a conductor. 

The magnetic fluctuations emanating from a conductor can be related to the material's conductivity $\sigma$ by first invoking the Biot-Savart law, which gives the magnetic field produced by current densities $J$ as $B_{z'}(\vec{r'}) = \mu_0 (J_{x'} y' - J_{y'} x') / (4\pi r'^3)$ where $\vec{r'} = (x',y',z')$ is the electron position and $\mu_0$ is the vacuum permeability, and then the Johnson-Nyquist formula, $S^{x',y',z'}_J(\omega) = 2 k_{\text{B}} T \text{Re}\left[\sigma(\omega)\right]$, where $S^{x',y',z'}_J$ is the spectral density of the current density fluctuations, $k_{\text{B}}$ is the Boltzmann constant, and $T$ is temperature. Note that $\sigma \equiv \text{Re}\left[\sigma(\omega)\right] \approx \sigma(0)$ for $\omega \sim 2 \pi * 2.87$ GHz. 

For the conductor geometry studied in this work, a thin-film of thickness $t_{\text{film}}$ a distance $d$ away from the NV sensor, a volume integral over the conductor yields the $z'$-component of the magnetic spectral density
\begin{flalign}
S_B^{z'} &= \frac{\mu_0^2 k_\text{B} T \sigma}{16 \pi} \left(\frac{1}{d} - \frac{1}{d + t_{\text{film}}}\right)
\label{integral}
\end{flalign}

and $S_B^{x',y'}=S_B^{z'}/2$ (a full derivation is given in supplementary information SI 1.1)
\cite{Henkel1999a,Langsjoen2014}. The spectral density component perpendicular to the NV axis is, for the $\left(1 0 0\right)$ cut diamond used in this work, $S_B^\perp = (4/3)S_B^{z'}$, which induces $|{m_{s}=0 \rightarrow \pm 1}\rangle$ transitions at a rate $\Omega_{\text{metal}}$. Applying perturbation theory to the NV center spin triplet (details in SI 1.3) yields the metal-induced relaxation rate

\begin{flalign}
\Gamma_{\text{metal}} = 3 \Omega_{\text{metal}} = \gamma^2 \frac{\mu_0^2 k_\text{B} T \sigma}{8 \pi} \left(\frac{1}{d} - \frac{1}{d + t_{\text{film}}}\right)
\label{gamma}
\end{flalign}

Thus a metal induces relaxation proportional to its conductivity $\sigma$. Further, the relaxing effect goes as $1/d$ for $d \ll t_{\text{film}}$ and $1/d^2$ for $d \gg t_{\text  {film}}$. In this work $t_{\text{film}} = 85$ nm and $d$ varies from 10 to 1000 nm, thus spanning both regimes.

The conductivity imaging setup consists of a laser scanning confocal microscope integrated with a tuning fork-based atomic force microscope (AFM) (Fig.~\ref{Setup}a) \cite{Pelliccione2014}. All experiments are performed in ambient conditions in a small applied magnetic field of $20$ G. The AFM scans an NV center within nanometer-scale proximity of the surface of a conducting sample and the confocal microscope is used to optically initialize and readout the spin state of the NV center. Optical access is through the 150-$\mu$m thick diamond plate. A waveguide patterned on the diamond is used to transmit microwaves to coherently drive transitions between the spin states. NV centers reside $\sim 7$ nm below the surface of the bulk piece of diamond and are formed by $^{14}$N implantation and subsequent annealing (details in Methods). To enhance photon collection efficiency, the diamond sample is patterned with 400-nm diameter nanopillars; only pillars containing 1 NV center are used here. Conducting samples are patterned onto custom-fabricated scanning probes with flat plateau-tips that have diameters of several micrometers (Fig.~\ref{Setup}a). The fabrication procedure (details in Methods) allows for a variety of sample geometries and materials, several of which we image in this work. The probes are then mounted onto a quartz tuning fork for AFM feedback and scanning.
To minimize relative position drift between the conducting sample and the NV center, we implement temperature stabilization to $\sim$ 1 mK/day in concert with active drift correction that utilizes AFM-based image registration.

\textbf{Electrical conductivity measurement.} To measure a metal's conductivity we measure the NV center relaxation rate $\Gamma_{\text{NV}} = 1 / T_1$ as a function of $d$ where

\begin{flalign}
\Gamma_{\text{NV}}\left(d,\sigma\right) = \Gamma_{\text{metal}}\left(d,\sigma\right) + \Gamma_{\text{NV,int}}
\label{NV}
\end{flalign}

where $\Gamma_{\text{NV,int}}$ is the intrinsic relaxation rate of the NV ($d  = \infty$), which is $\sim 200$ Hz for the NVs in this study. The $T_1$ is measured by initializing the NV into its $|{m_{s}=0}\rangle$ spin state with a 10-$\mu$s pulse of 532 nm light and then allowing the NV to decay for a dark time $\tau$ toward a thermally mixed state; this decay is measured via a subsequent spin-state-dependent photoluminescence (PL) measurement. For each $\tau$, two measurements are performed: in the first, we readout the PL of the NV $S(\tau)$; in the second, we insert a resonant microwave $\pi$ pulse after $\tau$ to swap the $|{0}\rangle$ and $|{-1}\rangle$ populations and then readout the PL $S_{\text{swap}}(\tau)$ (see SI Fig.~\ref{SCCPlot}a) \cite{Myers2017}. The difference $S - S_{\text{swap}}$ corresponds to the difference in population between the $|{0}\rangle$ and $|{-1}\rangle$ states, which decays to 0 with $\exp\left(-\tau/T_1\right)$ as plotted in Fig.~\ref{Setup}c (details in SI 1.2). The data in Fig.~\ref{Setup}c show a 5-fold reduction in NV $T_1$ in the presence of a metal film, corresponding to $\Gamma_{\text{metal}} = 840$ $\pm$ 60 Hz. 

\begin {figure}
 \includegraphics[width=90mm]{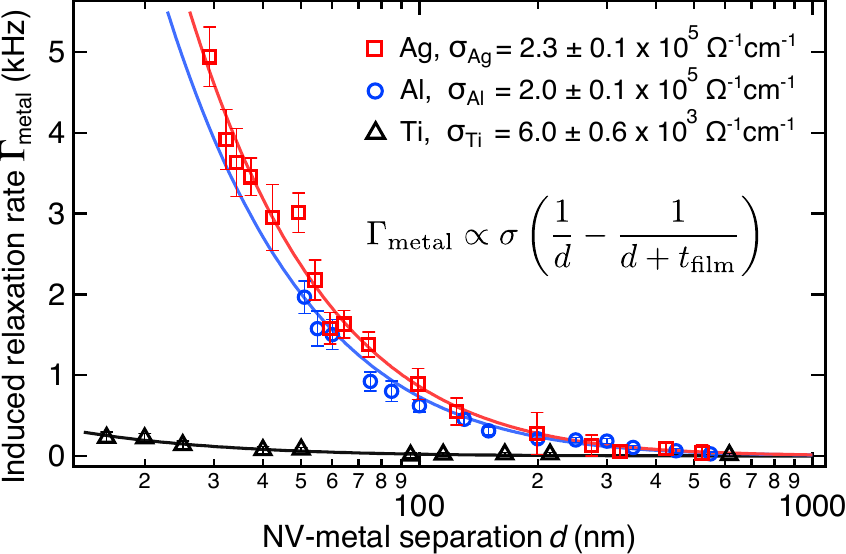}
\captionsetup{justification=raggedright, singlelinecheck=false}
\caption{Quantitative conductivity measurements using the NV center. Plotted is the metal-induced NV center relaxation rate $\Gamma_{\text{metal}}$ as a function of NV distance $d$ from 85-nm thick films of Ag (red squares), Al (blue circles), and Ti (black triangles). For each curve the intrinsic relaxation rate of the NV is measured and subtracted to isolate $\Gamma_{\text{metal}}$. The fits (solid curves) yield the conductivity values shown. Error bars correspond to the standard error in the fit to a full $T_1$ measurement at each point.}
\label{RelaxVsZ}
\end{figure}

We now demonstrate quantitative electrical conductivity measurements using the NV center. Figure~\ref{RelaxVsZ} plots $\Gamma_{\text{metal}}$ as a function of $d$ for three different 85-nm thick metal films of Ag, Al, and Ti, each measured with a different NV center. The films are deposited via thermal evaporation onto a $3$-$\mu$m diameter flat AFM tip as depicted in Fig.~\ref{Setup}a. Here, however, the film is continuous across the full extent of the plateau tip. Two qualitative observations can immediately be drawn from the data: first, $\Gamma_{\text{metal}}$ increases for small NV-metal separations; second, the highly conducting Ag and Al induce faster relaxation rates than Ti, a relatively poor conductor. These observations are consistent with Eq.~\ref{gamma}.

We quantitatively determine $\sigma$ by performing a least squares regression on the data in Fig.~\ref{RelaxVsZ} to Eq.~\ref{gamma}. The extracted values of $\sigma$ are $\sigma_{\text{Ag}} = 2.3 \pm 0.1 \times 10^{5}~ \Omega^{-1}\text{cm}^{-1}$, $\sigma_{\text{Al}} = 2.0 \pm 0.1 \times 10^{5} ~\Omega^{-1}\text{cm}^{-1}$, and $\sigma_{\text{Ti}} =  6.0 \pm 0.6 \times 10^{3} 
~\Omega^{-1}\text{cm}^{-1}$. These values are smaller than their bulk values by factors of 2.7, 1.7, and 4, respectively. A reduced conductivity is expected for metal films whose thickness is on the order of the electron mean free path ($\sim 50$ nm), and our measurements are consistent with experimental and theoretical values for 85-nm thick Ag, Al, and Ti films \cite{Lacy2011,Zhang2004}. We note that Eq.~\ref{gamma} assumes a slab of infinite extent, but here we study slabs with $\sim 3 ~ \mu$m diameter. However, this finite-size effect contributes deviations that are only $\sim$10\% of the experimental error and hence is negligible for this study (details in SI 1.4). 

In measuring $\Gamma_{\text{metal}}(d)$ we first contact the conducting sample to the diamond and then retract a known distance $d'$. The NV-metal separation in contact, $d_0$, is treated as a free fit parameter in the total NV-metal separation $d = d' + d_0$. Different values of $d_0$ are obtained for the three different NV-metal combinations, likely due to different tilt angles between the diamond and metal surfaces. 

We also note that the data in Fig.~\ref{RelaxVsZ} cannot be fit by a simple $1/d$ or $1/d^2$ dependence, as would be expected for $d\ll t_{\text{film}}$ or $d\gg t_{\text{film}}$, respectively, indicating that the NV's distance-dependent response is also sensitive to thickness of the conducting region.

\begin {figure*}
\centering
\includegraphics[width=146mm]{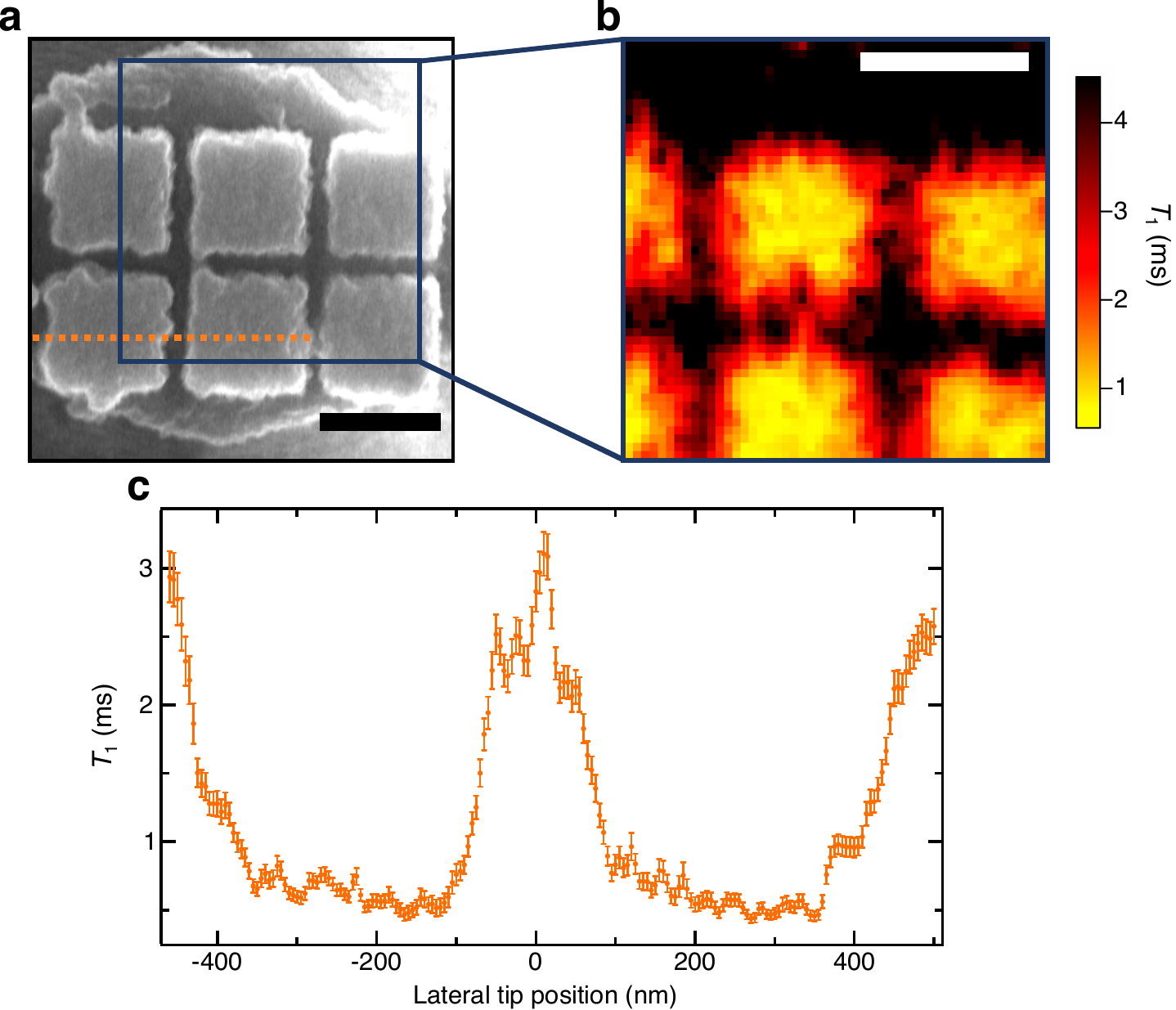}
\captionsetup{justification=raggedright, singlelinecheck=false}
\caption{Nanoscale conductivity imaging. \textbf{a} Scanning electron microscopy image of an Al nanopattern deposited onto an AFM tip, as depicted in Fig.~\ref{Setup}a. \textbf{b} NV $T_1$ image of the area depicted by the $1 ~\mu$m$^2$ blue square in \textbf{a}, produced by scanning the NV center over this area at a height of 40 nm and with 20 nm pixel spacing. \textbf{c} High-resolution $T_1$ line scan of the dotted orange line in \textbf{a}. Features in the imaging plane are clearly resolved down to a spatial resolution of 5 nm, which is set by the point spacing in the scan. This line scan implements a spin-to-charge measurement sequence, which results in experimental measurement error only 5x the spin projection noise limit and significantly reduced imaging time: 25x faster than standard spin-dependent NV photoluminescence readout. The intrinsic $T_1$ is 6 ms for both NVs. Scale bars are 400 nm. Error-weighted, light smoothing is applied to the data in \textbf{b} and \textbf{c}, for which nearest neighbors receive an additional weight reduction by a factor of 2.5. Error bars are calculated by propagating the measured standard error of the photoluminescence for the single-$\tau$ measurement of $T_1$.}
\label{CondImaging}
\end{figure*}

\textbf{Nanoscale conductivity imaging.} We now demonstrate nanoscale imaging of spatially inhomogeneous conductivity by laterally scanning an NV center over an array of  Al pads, as pictured in the scanning electron microscope (SEM) image in Fig.~\ref{CondImaging}a. This sample is formed by thermally evaporating 85 nm of Al onto 400 nm x 400 nm pads in silicon, fabricated by etching a grid of 350-nm deep, 100-nm wide trenches in the silicon. Figure~\ref{CondImaging}b plots the NV $T_1$ as the nanopatterned sample is scanned laterally over the NV. The $T_1$ is reduced when the NV is directly above the Al blocks and then recovers when above the gaps, clearly resolving the conducting features of our nanopatterned sample.

To expedite imaging we implement an adaptive, single-$\tau$ algorithm which sets the single $\tau$ point to be 0.7 $T_1$ of the previously measured pixel. The $\tau = 0$ point is also measured. This measurement method reduces the per-pixel measurement time to the order of a minute (details in SI 3). At these time scales, however, thermal drifts can still play a significant role and we perform active NV-sample drift correction via image registration. Topographic AFM images and NV PL images can both provide highly repeatable and sharp features for drift correction. In Fig.~\ref{CondImaging}b we use PL-based image registration every two hours to correct for $\sim$ 10 nm NV-sample drifts with $\sim$ 1 nm error (details in SI 2) \cite{Guizar-Sicairos2008}.

In Fig.~\ref{CondImaging}c we perform a high-resolution, 5-nm point spacing line scan of the dotted orange line in Fig.~\ref{CondImaging}a, demonstrating the nanoscale spatial resolution of our NV conductivity imaging technique. A different NV is used than for the image in Fig.~\ref{CondImaging}b.  Topographic-based image registration is performed once per hour. In addition to using an adaptive single-$\tau$ measurement, this measurement also implements a spin-to-charge readout sequence, which further reduces imaging time by a factor of 25 compared to standard spin-state dependent photoluminescence measurements and brings us to 5x the spin projection noise limit (details in SI 4). 

\subsection*{Discussion}

We now turn to a discussion of the spatial resolution and sensitivity of NV-based electrical conductivity imaging. From the line scan in Fig.~\ref{CondImaging}c, the metal-induced magnetic fluctuations at two points separated by 5 nm is resolved within the measured $T_1$ error, thus demonstrating 5 nm spatial resolution in the imaging plane. However, this does not necessarily correspond to the smallest resolvable conducting feature in the material. The ultimate resolution will also depend on AFM stability and NV-metal separation. The thermally-induced NV-sample drifts in Fig.~\ref{CondImaging} are $\sim$ 10 nm due to infrequent drift corrections, whose frequency could be increased to minimize thermal drifts, and ultimately, picometer-scale stability could be achieved with active drift compensation techniques \cite{King:2009gq}. For the data in Fig.~\ref{CondImaging}, the closest NV-metal separation $d_0$ is $\sim$ 40 nm, which sets a conductivity spatial resolution of $\sim$ 40 nm.

\begin {figure}
\includegraphics[width=88mm]{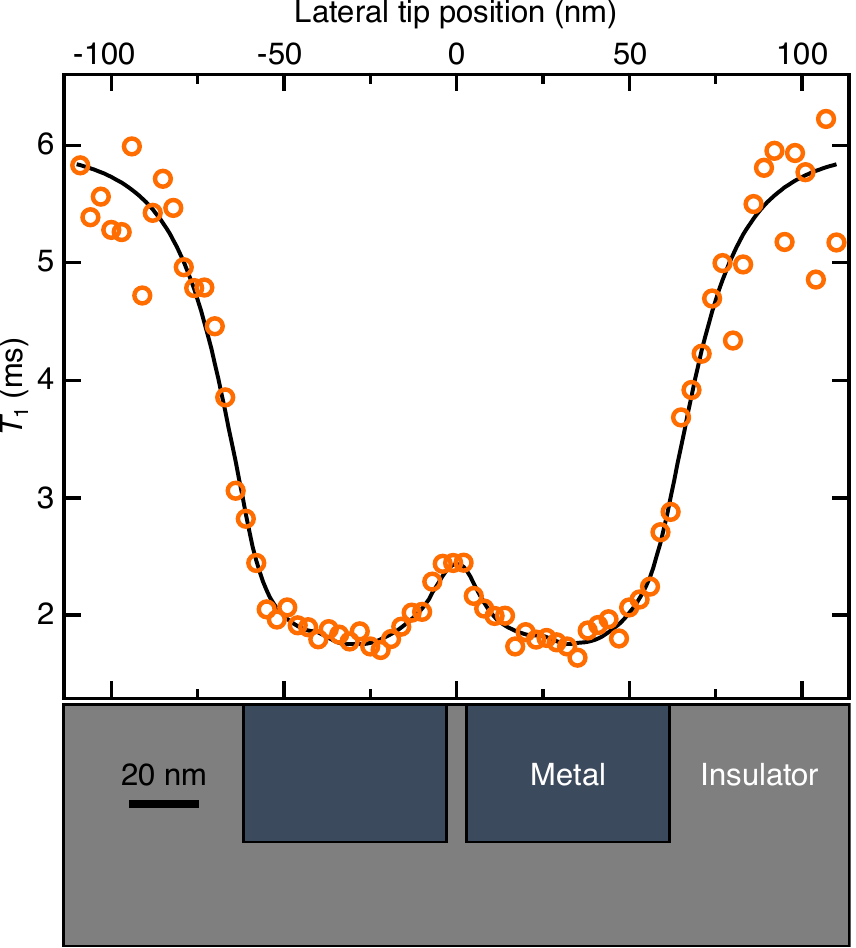}
\captionsetup{justification=raggedright, singlelinecheck=false}
\caption{Simulated NV $T_{1}$ line scan taken across a Mott insulator containing conducting pockets in an insulating material, as shown schematically at the bottom of the figure. The conducting regions, with $\sigma=3\times 10^{3} ~\Omega^{-1}\text{cm}^{-1}$, are  $40 \times 40 \times 60 ~\text{nm}^3$ in size and are separated by 5 nm. The black curve is the theoretical $T_1$ and the orange circles represent a simulated measurement using spin-to-charge readout with 1 minute of averaging per point.  The NV-metal separation is 5 nm and the intrinsic NV $T_1$ is 6 ms.}
\label{MITSimulation}
\end{figure}

For the mechanically and thermally stable imaging apparatus used here, the NV-metal separation limits both the spatial resolution and sensitivity of conductivity imaging. The smallest achievable separation is set by the NV depth in the diamond, $\sim$ 7 nm in this work. NV centers at few nm depths that exhibit several-ms $T_1$ times have also been measured \cite{Rosskopf:2014io}. The increased NV-metal separation we observe is likely dominated by an angular misalignment between the faces of the 3 $\mu$m diameter tip and the 400 nm diameter diamond pillar; a 5$\degree$ misalignment gives an NV-metal separation of $\sim$ 25 nm for an NV at the center of the pillar. For the Ti data in Fig.~\ref{RelaxVsZ} we measure $d_0 = 15 \pm 2$ nm, demonstrating the feasibility of close contact. With more controlled tilting and shallower NV centers, a 5-nm NV-metal separation should be achievable.

An important advance presented here is the implementation of NV spin-to-charge conversion (SCC) readout techniques \cite{Shields:2015ik} in imaging with shallow NV centers. In doing so we significantly reduce imaging time by a factor of 25, which is particularly relevant for relaxation imaging, an inherently long measurement due to the ms-scale $T_1$ times and point-by-point scanning. Notably, we find that the SCC readout technique is highly robust for shallow NV centers; all measured photostable NVs exhibit a significant enhancement in sensitivity, with a typical 20-30x reduction in measurement time (details in SI 4). Thus SCC should find ubiquitous utility for shallow NV sensing and imaging. By enhancing sensitivity, SCC techniques also extend the measurable range of conductivities and hence the range of accessible condensed matter phenomena such as phase separation in complex oxides \cite{Qazilbash:2007cm}, as we show next.

To illustrate the feasibility of resolving spatial conductivity variations in a relevant material system, in Fig.~\ref{MITSimulation} we plot a simulated $T_1$ line scan for an NV center scanned across a material with pockets of metallic phases inside an insulating matrix, as is seen in complex oxides across a metal-to-insulator transition \cite{Yee:2015jn, Qazilbash:2007cm}. The two nanoscale metallic regions and the 5-nm wide insulating barrier separating them are clearly resolved, both in the theoretical $T_1$ curve (black line) and in the simulated measurement (orange circles). The simulated measurement accounts for the expected measurement error with 1 minute of averaging per point, an adaptive-$\tau$ SCC readout technique, $d_0$ = 5 nm, an intrinsic NV $T_1$ of 6 ms, and $\sigma_{\text{metal}}=3\times 10^{3}~ \Omega^{-1}\text{cm}^{-1}$. This conductivity value is typical for a Mott insulator \cite{Imada1998}, and is only half the conductivity of Ti measured in Fig.~\ref{RelaxVsZ}. To simulate the magnetic fluctuations from the conducting pockets we implement a Monte Carlo simulation of the electron trajectories (details in SI 1.4) \cite{Bulashenko1992}. This simulation demonstrates the feasibility of noninvasive, nanoscale, NV-based imaging of inhomogeneous electrical conductivity in Mott insulators. 

The NV center in diamond is emerging as a versatile quantum sensor capable of imaging magnetism \cite{Rondin2013,Maletinsky2012}, temperature \cite{Kucsko2013}, thermal conductivity \cite{Laraoui2015}, and DC currents \cite{Tetienne2017,Chang2017}. In this paper we add conductivity to the arsenal of NV imaging modalities. Future studies can, for example, combine DC magnetic field and conductivity sensing in one integrated tool to yield unique insight into materials with multiple order parameters, such as Nd$_2$Ir$_2$O$_7$ with the possibility of domain wall conductivity at magnetic domain walls \cite{Ma2015}, and buried LAO/STO interfaces with coexisting superconductivity and ferromagnetism \cite{Bert2011}.

\subsection*{Methods}
\textbf{Experimental setup and sample preparation.} The experimental setup consists of a home-built confocal microscope and atomic force microscope. The experiments are done in ambient conditions with active temperature control to within 1 mK. A continuous wave laser at 532 nm is used for optical pumping and readout of the NV spin, and is gated with an acousto-optic modulator (AOM). For spin-to-charge readout, continuous wave lasers at 594 nm and 637 nm gated with AOMs are also used. Photons emitted by the NV are collected into a single-mode fiber and directed to a fiber-coupled avalanche photodiode. The collection efficiency is amplified by a factor of $\sim$5 due to waveguiding from the 400-nm wide, 500-nm tall, truncated-cone diamond pillars. Microwaves used for resonant spin manipulation are delivered via a 300-nm thick waveguide evaporated onto the diamond surface closest to the NVs. Timing of the pulse sequences is controlled by a Spincore Pulseblaster ESR-Pro 500~ MHz card.

Nanopatterned metal samples are fabricated in a top-down process starting with Silicon on Insulator wafers with a 10-$\mu$m thick device layer. Nanopatterns are defined in the device Si layer using electron beam lithography and reactive-ion etching. Subsequently, a standard AFM probe fabrication process flow is carried out. The flat-faced tip is formed by stopping the KOH anisotropic Si etch such that the apex of the Si pillar retains a $\sim$2 $\mu$m diameter with the nanopattern untouched. Cantilevers are released via a backside Si deep-etch and an HF SiO$_2$ etch. 85 nm of metal is then thermally evaporated onto the tip face.

To perform AFM, these cantilevers are then glued onto tuning forks using a micromanipulator. To minimize angular misalignments between the tip face and the diamond surface, we first use an SEM to measure the relative tilt between the tip face and the tuning fork mount. We then tilt the entire tuning fork assembly with respect to the diamond to make them parallel to within several degrees, which is limited by the SEM resolution of nonconducting silicon. Existing SEM technologies specifically designed for semiconductors could improve this resolution and tilt adjustment. We operate in tapping mode AFM, in which we electrically drive the quartz tuning fork to amplitudes of $\sim$ 1 nm, and we measure the amplitude with a lock-in amplifier for feedback control (Zurich Instruments). XYZ positioning is controlled by piezoelectric scanners.

The diamond is prepared via growth of a 50-nm thick 99.99\% $^{12}$C isotopically purified thin film on a commercial Element 6 electronic grade (100) diamond substrate. Prior to growth, the diamond is etched with ArCl$_2$ plasma (1 $\mu$m) to mitigate polishing damage and cleaned in boiling acid H$_2$NO$_3$:H$_2$SO$_4$. NV centers are formed by $^{14}$N ion implantation with a dosage of $5.2\times 10^{10}$  ions / cm$^2$ at 4  keV and a 7$\degree$ tilt, which yields an expected depth of 7  nm (calculated by Stopping and Range of Ions in Matter (SRIM)). The sample is then annealed in vacuum ($< 10^{-6}$ Torr at max temperature) at 850$\degree$ C for 2.5 hours with a 40-minute temperature ramp. After annealing, the sample is cleaned in HClO$_4$:H$_2$NO$_3$:H$_2$SO$_4$ 1:1:1 for 1 hour at 230-240 $\degree$C.

\textbf{Error analysis.} Errors in measured $T_1$ are given by the standard error in the exponential fit where the decay constant $T_1$ and amplitude are the only free parameters, as in Fig.~\ref{Setup}c. In Fig.~\ref{RelaxVsZ} a full $T_1$ measurement is done at each point and error bars correspond to the standard error in the fit. In the case of the single-$\tau$ measurement in Fig.~\ref{CondImaging}, $T_1$ is explicitly calculated and the error results from propagating the measured standard error of the photoluminescence. Error in the measured conductivities is given as the standard error in the fit to the data in Fig.~\ref{RelaxVsZ}.

\textbf{Data availability.} All relevant data are available upon request from A.C.B.J.

\subsection*{Acknowledgements}
We thank Shimon Kolkowitz and Nathalie de Leon for helpful discussions. We also thank Matthew Pelliccione for constructing the scanning probe setup. This work was supported by a PECASE award from the Air Force Office of Scientific Research. A.A. acknowledges funding from the Elings Postdoctoral Prize Fellowship from the UCSB California NanoSystems Institute. 
 
\subsection*{Author contributions}
A.A. fabricated the samples. A.A. and D.B. performed the experiments and data analysis. B.A.M. developed most of the experiment infrastructure. A.A., D.B., and A.C.B.J wrote the paper. A.C.B.J. supervised the project. All authors contributed to the design of the experiment and discussions during the course of the measurements and analysis.

\subsection*{Additional information}
\textbf{Competing interests:} The authors declare no competing financial interests.

\bibliography{biblio}

\clearpage
\onecolumngrid
\subsection*{\large Supplemental Information}
\normalsize
\maketitle
\setcounter{equation}{0}
\setcounter{figure}{0}
\setcounter{table}{0}
\setcounter{page}{1}
\makeatletter
\renewcommand{\theequation}{S\arabic{equation}}
\renewcommand{\thefigure}{S\arabic{figure}}

\section*{SUPPLEMENTARY NOTE 1: THEORETICAL MODEL OF NV RELAXATION}
\subsection*{1.1: Magnetic spectral density emanated by a thin conducting film}

To quantitatively measure a material's conductivity using an NV center in diamond, we use theory developed by \cite{Henkel1999a,Langsjoen2014} to describe the magnetic Johnson Noise produced by an infinite conducting slab of finite thickness $a$. The experiments presented here probe polycrystalline metals at room temperature, and therefore we do not consider non-local 
effects such as those observed in \cite{Kolkowitz2015}. We then develop an analytic solution to the magnetic spectral density from a conductor with finite thickness $a$. For consistency with referenced works, in this supplement we define the metal thickness as $a$ and the distance of the NV to the conductor surface as $z$, instead of $t_{\text{film}}$ and $d$, as used in the main text, respectively.

First, consider the commonly presented solution as in \cite{Henkel1999a,Kolkowitz2015} for the magnetic spectral density emanated by a metallic half-space

\begin{equation}
\begin{split}
\label{half-space}
S^{z'}_{B,\text{half-space}} &= \frac{\mu_0^2 k_B T \sigma}{16 \pi} \left( \frac{1}{z} \right)
\\
S_{B,\text{half-space}}^{x'} & = S_{B,\text{half-space}}^{y'} = \frac{S_{B,\text{half-space}}^{z'}}{2}
\end{split}
\end{equation} 

Where $\mu_0$ is the vacuum permeability, $k_{\text{B}}$ the Boltzmann constant, $T$ the temperature, and $\sigma$ the metal conductivity.
A half-space a distance $z$ away is equivalent to the summation of a slab of thickness $a$ a distance $z$ away and a half-space a distance $z + a$ away.

\begin{equation}
\begin{split}
&S_{B,\text{half-space}}^{z'}(z) = S_{B,\text{slab}}^{z'}(z) + S_{B,\text{half-space}}^{z'+a}(z+a)
\\
&S_{B,\text{slab}}^{z'}(z) = S_{B,\text{half-space}}^{z'}(z) - S_{B,\text{half-space}}^{z'+a}(z+a)
\\
&S_{B,\text{slab}}^{z'} = \frac{\mu_0^2 k_B T \sigma}{16 \pi} \left( \frac{1}{z} - \frac{1}{z+a} \right)
\label{geometric}
\end{split}
\end{equation}

This elegant solution neglects the effect of the boundary at $z+a$. To confirm its validity, we rigorously derive the same result using Fresnel coefficients for a finite-thickness film.

Derived using the fluctuation-dissipation theorem and a magnetic Green's tensor, the magnetic spectral density tensor at angular frequency $\omega$ and temperature $T$ can be expressed as 

\begin{equation}
S_B^{ij} = \frac{\hbar \omega^3}{4 \pi \epsilon_0 c^5} \coth{\left(\frac{\hbar \omega}{2 k_B T}\right)} s_{ij}
\end{equation}

Where $\hbar \omega^3 / 4 \pi \epsilon_0 c^5$ has units of magnetic spectral density (Tesla$^2$ / Hz), and $s_{ij}$ is a dimensionless tensor. For simplicity we consider a coordinate system $x',y',z'$ with the $z'$-axis perpendicular to the metal, and $z$ being the distance to the nearest face of the metal film, such that $s_{ij}$ becomes a diagonal tensor with elements 

\begin{equation}
\begin{split}
\label{integrals}
s_{x'x'} = s_{y'y'} & = \frac{1}{2} \text{Re} \int_0^{\infty} du \hspace{1.5pt} \frac{u \left(r_p(u) + (u^2 - 1)r_s(u) \right)}{\eta} e^{2izk\eta} \\
s_{z'z'} & = 1 \hspace{2pt} \text{Re} \int_0^{\infty} du \hspace{1.5pt} \frac{u^3 r_s(u)}{\eta} e^{2izk\eta}
\end{split}
\end{equation}

\begin{equation}
\eta(u) = \begin{cases}
\sqrt{1 - u^2} & 0 \leq u \leq 1 \\
i \sqrt{u^2 - 1} & u > 1
\end{cases}
\end{equation}

Where we use $k = \left|\omega\right|/c$ and Fresnel coefficients $r_p$ and $r_s$. For simplicity we drop the double index on the diagonal elements such that $s_{z'} \equiv s_{z'z'}$
As shown in \cite{Langsjoen2014}, for a metal of finite thickness $a$ these Fresnel coefficients are

\begin{align}
\label{fresnel}
r_s(u) = \frac{k_1^2 - k_2^2}{k_1^2 + k_2^2 + 2 i k_1 k_2 \cot{\left(k_2 a\right)}} && r_p(u) = \frac{(\epsilon_2 k_1)^2 - (\epsilon_1 k_2)^2}{(\epsilon_2 k_1)^2 + (\epsilon_1 k_2)^2 + 2 i \epsilon_1 \epsilon_2 k_1 k_2 \cot{\left(k_2 a\right)}}
\end{align}

\begin{align}
k_1 \equiv k \sqrt{\epsilon_1 - u^2} && k_2 \equiv k \sqrt{\epsilon_2 - u^2}
\label{K_eqn}
\end{align}

Where $\epsilon_{1,2}(\omega)$ are the relative dielectric functions of the two media, which in this case are $\epsilon_1 \equiv \epsilon_{\text{vacuum}} = 1$ and $\epsilon_2 \equiv \epsilon_{\text{metal}}$. 
In the frequency regime of the NV level splitting $\omega \approx $ $2\pi *$ 2.8 GHz, and for metal conductivity $\sigma$

\begin{equation}
\epsilon_2 \approx \frac{i \sigma}{\epsilon_0 \omega}
\end{equation}

Thus, the above expressions relate the magnetic noise spectral density at the NV produced by a metal with conductivity $\sigma$.
 
As described in \cite{Henkel1999a} the integrals are dominated by values of $u \approx 1/(2kz) \approx 10^6$ for our experiment where $z \approx 10$ nm and $\omega \approx 2\pi * 2.8$ GHz. For the metallic conductivities studied here, $|\epsilon_2|^{1/2}\approx 10^3 $ - $ 10^4$, and thus $u \gg |\epsilon_2|^{1/2}$. This is equivalent to stating that the skin depth $\delta \gg z$, where $|\epsilon_2|^{1/2} = \sqrt{2} / k \delta$. We now make the critical approximation that $\sqrt{\epsilon - u^2} \approx \sqrt{1 - u^2} \approx i u$ and set $\epsilon_1 = 1$ for vacuum

\begin{equation}
r_s(u) = \frac{1 - u^2 - \epsilon_2 + u^2}{1 - u^2 + \epsilon_2 - u^2 + 2 i \sqrt{1-u^2}\sqrt{\epsilon_2 - u^2} \cot{\left(k a \sqrt{\epsilon_2 - u^2}\right)}} = \frac{1 - \epsilon_2}{1 + \epsilon_2 - 2u^2 ( 1 + \coth{(kau)})}
\end{equation}

Multiplying the numerator and denominator by the complex conjugate of the denominator

\begin{equation}
r_s(u) = \frac{\left(1 - \epsilon_2\right)\left(-\epsilon_2 - 2u^2(1 + \coth(kau))\right)}{|\epsilon_2|^2 + 4u^4(1 + \coth(kau))^2}
\end{equation}

Applying $\sqrt{1 - u^2} \approx i u$ to the $s_{z'}$ integral in Eq.~\ref{integrals} yields

\begin{equation}
\begin{split}
s_{z'} &= 1 \hspace{2pt} \text{Re} \int_0^{\infty} du \hspace{1.5pt} \frac{u^3}{i u} r_s(u) e^{2izk (i u)} \\
&= 1 \hspace{2pt} \text{Re} \int_0^{\infty} du \hspace{1.5pt} \frac{u^2}{i} \frac{\left(1 - \epsilon_2\right)\left(-\epsilon_2 - 2u^2(1 + \coth(kau))\right)}{|\epsilon_2|^2 + 4u^4(1 + \coth(kau))^2} e^{-2kzu}
\label{coth}
\end{split}
\end{equation}

$\coth{(kau)} + 1$ is a real, positive number always greater than 2, thus $4u^4 (1 + \coth(kau))^2 \gg |\epsilon_2|^2$. Taking the real part,

\begin{equation}
s_{z'} = \frac{\epsilon_2}{i} \int_0^{\infty} du \frac{2u^4 (1 + \coth(kau))}{4u^4\left(1 + \coth(kau)\right)^2} e^{-2kzu}
\end{equation}

Noting that $1 + \coth(kau) = 2e^{kau} / (e^{kau} - e^{-kau})$, and substituting in $\epsilon_2 = i \sigma / (\epsilon_0 \omega)$

\begin{equation}
\begin{split}
s_{z'} & = \frac{\sigma}{\epsilon_0 \omega} \int_0^{\infty} du  \frac{(e^{-2kzu})(e^{kau} - e^{-kau})}{4e^{kau}}
\\
& = \frac{\sigma c }{8 \epsilon_0 \omega^2} \left( \frac{1}{z} - \frac{1}{z+a} \right)
\end{split}
\end{equation}

Using the same approximations as above one finds $r_p(u) \approx 1$, and then that $s_{x'} = s_{y'} \approx s_{z'} /2$. Further, at room temperature and $\omega \approx$ $2\pi *$ 2.8 GHz, $\coth{(\hbar \omega / 2 k_B T)} \approx 2 k_B T / \hbar \omega$. Calculation of the dimensionless tensor $s_{ij}$ then yields the magnetic spectral density tensor elements

\begin{equation}
\begin{split}
S_B^{z'} & = \frac{\hbar \omega^3}{4 \pi \epsilon_0 c^5} \frac{2 k_B T}{\hbar \omega} \frac{\sigma c}{8 \epsilon_0 \omega^2} \left( \frac{1}{z} - \frac{1}{z+a} \right) 
\\
S_B^{z'} & = \frac{\mu_0^2 k_B T \sigma}{16 \pi} \left( \frac{1}{z} - \frac{1}{z+a} \right)
\\
S_B^{x'} & = S_B^{y'} = \frac{S_B^{z'}}{2}
\label{SBz}
\end{split}
\end{equation}

As expected, Eq.~\ref{SBz} simplifies to the half-space solution as one takes film thickness $a \rightarrow \infty$, and the calculation performed here matches our geometric derivation in Eq.~\ref{geometric}. Importantly, the Johnson noise is white for the relevant GHz frequencies.

We now complete this analysis by relating the magnetic spectral density to the ensuing relaxation rate from $\ket{m_s = 0}$ to $\ket{m_s = -1}$ for a spin-1 system like the NV.

\subsection*{1.2: Population dynamics of the NV three-level system}

As described in \cite{Myers2017}, the relaxation rate of $\ket{m_s = 0}$ to $\ket{m_s = 1}$ is approximately the same as the relaxation rate of $\ket{m_s = 0}$ to $\ket{m_s = -1}$. Denoting this rate as $\Omega$, and denoting the double-quantum transition rate between $\ket{m_s = 1}$ and $\ket{m_s = -1}$ as $\gamma$, the population dynamics are thus described by

\begin{equation}
\frac{d}{dt} 
\begin{pmatrix}
	\rho_0 \\
    \rho_{-1} \\
    \rho_1
\end{pmatrix}
=
\begin{pmatrix}
    -2\Omega & \Omega & \Omega \\
    \Omega & -\Omega - \gamma & \gamma \\
    \Omega & \gamma & -\Omega - \gamma
\end{pmatrix}
\begin{pmatrix}
	\rho_0 \\
    \rho_{-1} \\
    \rho_1
\end{pmatrix}
\end{equation}

Solving the eigenvalue-eigenvector problem, we see that

\begin{equation}
\begin{pmatrix}
	\rho_0 \\
    \rho_{-1} \\
    \rho_1
\end{pmatrix}
=
C_1
\begin{pmatrix}
	2 \\
    -1 \\
    -1
\end{pmatrix}
e^{-3 \Omega t}
+ 
C_2
\begin{pmatrix}
	0 \\
    1 \\
    -1
\end{pmatrix}
e^{-\left(\Omega + 2 \gamma\right) t}
+ 
C_3
\begin{pmatrix}
	1\\
    1\\
    1
\end{pmatrix}
\end{equation}

As described in the main text, we optically polarize the NV into $\ket{m_s = 0}$. Ideally $\rho_0(0) = 1$ and $\rho_{-1}(0) = \rho_{1}(0) = 0$, but we account for imperfect optical polarization by defining $\eta$ such that $\rho_0(0) = 1 - 2 \eta$ and $\rho_{-1}(0) = \rho_{1}(0) = \eta$, where $\eta$ is nominally 0.05 \cite{Shields:2015ik}. With these initial conditions, we see that

\begin{align}
\rho_0(t) &= \frac{1}{3} + 2 \left(\frac{1}{3} - \eta\right) e^{-3 \Omega t}
\\
\rho_{-1}(t) = \rho_1(t) &= \frac{1}{3} - 1 \left(\frac{1}{3} - \eta\right) e^{-3 \Omega t}
\end{align}

In a spin-dependent photoluminescence (SDPL) measurement sequence, such as shown in Fig.~\ref{SCCPlot}a, optical polarization is followed by evolution for a dark time $\tau$, and then the NV photoluminescence (PL) signal $S$ is measured, where $S$ is given by

\begin{equation}
S = A \rho_0  + B \rho_{-1} + B \rho_1 + \text{Background}
\end{equation}

$A$ and $B$ describe the brightness of the $\ket{0}$ and $\ket{\pm1}$ states, respectively. The $\ket{-1}$ and $\ket{1}$ states are taken to be equally bright. Background designates any non-NV$^-$ signal that may come from \textit{e.g.} APD dark counts, light leakage, and any population in the neutral NV$^0$ due to imperfect charge state polarization.
We then repeat this experiment, but immediately before readout a resonant microwave pulse swaps the population of the $\ket{-1}$ and $\ket{0}$ states. This second measurements yields the signal $S_{\text{swap}}$:

\begin{equation}
S_{\text{swap}} = B \rho_0 + A \rho_{-1} + B \rho_1 + \text{Background}
\end{equation}

We then take the difference between the two measurements $S$ and $S_{swap}$ to calculate the quantity $S_{\text{diff}}$

\begin{equation}
S_{\text{diff}} = S - S_{\text{swap}} = \left(A - B \right) \left(1 - 3 \eta\right) e^{-3 \Omega t} = C e^{-3 \Omega t}
\label{Seqn}
\end{equation}

where we include the imperfect spin polarization in the definition of contrast $C$. In Fig.~\ref{Setup}c of the main text we plot a normalized version of $S_{\text{diff}}$, where the contrast is normalized to 1 by means of reference measurements of $S_{\text{diff}}(0)$. We fit to an exponential decay with decay rate

\begin{equation}
\frac{1}{T_1} = \Gamma = 3 \Omega
\end{equation}

Double-quantum relaxation, imperfect spin polarization, and any background signal do not modify the measured $\Gamma$. However, AOM leakage during long, ms-scale dark times, can repolarize the NV at a rate $\delta$, which changes the form of the decay curve to $S_{\text{diff}}=C \exp\left(-\left(3\Omega + \delta\right)t\right) + y_0$. $y_0$ is an inevitable offset since the dark, steady-state population with a slow polarization rate will not be an even thermal mixture of the 3 spin states. We find that several nW of laser leakage can repolarize the NV at a rate of $\sim$ 50 Hz. Thus we take care to reduce laser leakage to below 1 nW and confirm that $S_{\text{diff}} \rightarrow 0$ as $t \rightarrow \infty$.

\subsection*{1.3: NV relaxation rate near a conductor}

We now calculate $\Omega_{\text{metal}}$ via perturbation theory applied to the NV Hamiltonian, with NV coordinates $x,y,z$. The NV's magnetic moment axis, the $z$-axis, makes an angle $\theta$ with the $z'$-axis. We can further select this axis to lay entirely in the $x'$-$z'$ plane without any loss of generality, since the $x'$ and $y'$ components of the magnetic spectral density tensor are equivalent.

Hence, $B_{x} =  \cos(\theta) B_{x'} + \sin(\theta) B_{z'}$ and $B_{y} = B_{y'}$. Note that the $x',y',z'$ components of the magnetic field are all uncorrelated, and thus $S_B^{x} = \cos^2{(\theta)} S_B^{x'} + \sin^2{(\theta)} S_B^{z'}$ and $S_B^{y} = S_B^{y'}$.
We use the formalism described in \cite{Henkel1999a} to describe the relaxation rate under the interaction Hamiltonian $H'$, as described in the main text with NV gyromagnetic ratio $\gamma$ and raising and lowering operators $S_+$ and $S_-$

\begin{flalign}
H' = \gamma B_{z}S_{z} + \frac{1}{2}\gamma(B_{x} - iB_{y})S_{+} + \frac{1}{2}\gamma(B_{x} + iB_{y})S_{-}
\label{hamiltonian2}
\end{flalign}

\begin{equation}
\Omega_{0 \rightarrow -1, \text{metal}} = |\braket{0|\gamma S_z|-1}|^2 S_B^{z} + |\braket{0|\frac{1}{2}\gamma S_+|-1}|^2 \left(S_B^{x} + S_B^{y}\right) + |\braket{0|\frac{1}{2}\gamma S_-|-1}|^2 \left(S_B^{x} + S_B^{y}\right)
\end{equation}

\begin{equation}
\Gamma_{\text{metal}} = 3 \Omega_{\text{metal}} = 3 \gamma^2 \left(\frac{1}{2} \cos^2{(\theta)}S_B^{x'} + \frac{1}{2} S_B^{y'} + \frac{1}{2} \sin^2{(\theta)} S_B^{z'} \right)
\end{equation}

The diamonds used in this work are $(100)$ oriented, and thus all four possible NV orientations make an angle $\theta = \arccos(\sqrt{1/3}) \approx 54.7^{\circ}$ with the $z'$-axis, making our analysis independent of the NV orientation. 

We now finish our derivation of relaxation rate by using our derived expression for the magnetic spectral density tensor.

\begin{equation}
\Gamma_{\text{metal}} = \gamma^2 \frac{3 \mu_0^2 k_B T \sigma}{32 \pi} \left(1 + \frac{1}{2} \sin^2{(\theta)} \right)\left(\frac{1}{z} - \frac{1}{z+a}\right)
\end{equation}

\begin{equation}
\Gamma_{\text{metal}} = \gamma^2 \frac{\mu_0^2 k_B T \sigma}{8 \pi} \left(\frac{1}{z} - \frac{1}{z+a}\right)
\end{equation}

Note that in the main text we define $d \equiv z$ and $t_{\text{film}} \equiv a$.

\subsection*{1.4: Calculation of the magnetic spectral density emanated by a finite-geometry conductor}

The magnetic fluctuations emanating from a conductor can be derived in an alternate method to that presented in section 1.1. This method is the one presented in the main text and it offers much physical intuition. However, we warn that the results for the $x'$ and $y'$ components of the magnetic spectral density are overestimated by a factor of 3 because we do not account for boundary conditions at the surface. Nevertheless, this method has further utility as it allows us to estimate the magnetic noise from finite geometries \cite{Henkel2005}.

In the Drude model, the $i^{th}$ component ($i = x',y',z'$) of an electron's velocity $v_i$ will be correlated in time by $\langle v_i(t)v_i(t+t')\rangle = v_i^2 * \exp(-t/\tau_c)$, where $\tau_c$ is the mean electron collision time. With the Wiener-Khinchin theorem, taking a Fourier transform of this velocity autocorrelation function yields a two-sided velocity spectral density

\begin{flalign}
{S_v^{i}}(\omega) = {v_i^2} \frac{\tau_c}{1+\omega^2 \tau_c^2}
\label{vel}
\end{flalign}

The velocity noise produced by an electron with mass $m$ and mean thermal energy $3k_\text{B} T/2$ is thus

\begin{flalign}
\overline{S_v^{i}}(\omega) = \frac{2 k_\text{B} T}{m} \frac{\tau_c}{1+\omega^2 \tau_c^2}
\end{flalign}

The Biot-Savart law gives the magnetic field from a moving electron as $B_{z'}(t) = \mu_0 (v_{y'}(t) x' - v_{x'}(t) y') e / 4 \pi r'^3$, where $e$ is the electronic charge and $r'$ is the distance to the electron. From the autocorrelation function $\langle B_{z'}(t)B_{z'}(t+t')\rangle$ we calculate $S_{B,\text{one electron}}^{z'} = \mu_0^2 e^2 (y'^2 S_v^{x'} + x'^2 S_v^{y'}) / (4 \pi r'^3)^2$. Note that using the Wiener-Khinchin theorem to find the velocity spectral density is equivalent to the argument in the main text where we start at the Johnson-Nyquist expression for current spectral density.

We assume all of the electrons in the metal are uncorrelated, and thus their magnetic field spectral densities add incoherently. Summing up the mean electron contributions from volume elements with electron density $n$, we then calculate $S_B^{z'}$ from an infinite film of thickness $a$ a distance $z$ away,

\begin{flalign}
S_B^{z'} &= \frac{\mu_0^2 k_\text{B} T}{16 \pi} \left(\frac{n e^2}{m}\frac{\tau_c}{1+\omega^2\tau_c^2}\right)\int_{V'} dV' \frac{4\left(x'^2 + y'^2\right)}{2 \pi r'^6}
\\
S_B^{z'} &= \frac{\mu_0^2 k_\text{B} T}{16 \pi}\sigma \int_z^{z+a} \int_0^{\infty} \frac{4 \rho'^3 d\rho' dz'}{\left(\rho'^2 + z'^2\right)^3}
\label{integralSup}
\\
S_B^{z'} &= \frac{\mu_0^2 k_\text{B} T \sigma}{16 \pi} \left(\frac{1}{z} - \frac{1}{z + a}\right)
\end{flalign}

where $\sigma$ is the electrical conductivity with $\sigma \equiv \text{Re}\left[\sigma(\omega)\right] \approx \sigma(0)$ for $\omega \sim 2 \pi * 2.87$ GHz. This calculation method yields the correct expression for $S_B^{z'}$, which was rigorously derived in section 1.1 using the fluctuation-dissipation theorem. However, because we did not account for boundary conditions at the surface $S_B^{x'}$ and $S_B^{y'}$ would be overstated by a factor of 3 with this method of calculation \cite{Henkel2005}.

In Fig.~\ref{RelaxVsZ} of the main text we study the noise from plateau tips of $1.5~ \mu$m radius, which we approximate to be infinite films. We can conveniently estimate the deviation from the infinite-film approximation by performing the radial integral in Eq.~\ref{integralSup} to 1.5 $\mu$m, instead of $\infty$. Similarly, we calculate integral expressions for $S_B^{x'}$ and $S_B^{y'}$ and integrate to 1.5 $\mu$m. We then compare $\Gamma_{\text{Biot},\infty}$ to $\Gamma_{\text{Biot,1.5$\mu$m}}$. Although the simplistic Biot-Savart calculation ignores surfaces and thus is slightly skewed in magnitude, by considering the volume integral we closely estimate the relative deviation from the infinite-film model. We find that this relative deviation is approximately 10\% of the experimental error for $\Gamma_{\text{metal}}$ for Fig.~\ref{RelaxVsZ}, which we deem to be negligible for this study.

This Biot-Savart method of calculation also allows us to form the basis for the simulation in Fig.~\ref{MITSimulation} of the main text. We employ a Monte Carlo simulation of a single electron in a metal, which has previously been developed to estimate the velocity autocorrelation function and spectral density inside finite geometries \cite{Bulashenko1992}. We simulate an electron with a given velocity and a certain probability to scatter dictated by the bulk, mean collision time $\tau_c$. The electron also scatters at boundaries, however, and this effect modifies the velocity spectral density: near the boundary the effective collision time $\tau_c$ will decrease, which stretches the velocity spectral density (Eq.~\ref{vel}) and suppresses the magnitude of the noise at $\omega \sim 2 \pi * 2.87$GHz.

For the simulation in Fig.~\ref{MITSimulation} of the main text, we perform a Monte Carlo simulation of the electron trajectory, but instead of calculating the velocity noise we use the Biot-Savart law to explicitly calculate the magnetic autocorrelation spectral density for every point in free space. This simulation allows us to account for the finite size of the conductor as well as the approximate noise suppression at boundaries. Since the noise from two conductors will be uncorrelated, we can sum the spectral densities from the two conducting blocks in Fig.~\ref{MITSimulation}. This can be done for any geometry. In order to ensure that the magnitude of the magnetic spectral density is properly estimated, we employ the same logic as for the estimation described above: we compare the simulated $\Gamma_{\text{sim, finite}}$ to $\Gamma_{\text{sim},\infty}$ to estimate the relative deviation from an infinite film and we then apply Eq.~\ref{SBz} to obtain the absolute magnitude.

\section*{SUPPLEMENTARY NOTE 2: TEMPERATURE CONTROL AND SPM STABILITY}

Temperature fluctuations play an important role in our system's stability; a 1 mK change results in $\sim$ 1 nm of drift of the SPM cantilever relative to the NV, as a consequence of using materials with different coefficients of thermal expansion. To mitigate thermal drifts, we implement several layers of thermal isolation and active temperature feedback. Our experimental apparatus is enclosed in an insulating box on an optical table, with the entire optical table isolated by curtains. The laboratory temperature is stabilized to within 1K. A PID loop measures voltage across a thermistor with a sensitivity 0.1 mK / Hz$^{1/2}$ and then heats thin resistance wire in feedback control. Thin resistance wire is optimal, with a small thermal mass and large surface area relative to its volume, yielding responsive feedback. This also allows us to distribute the heating sources evenly around the insulating box, which is critical in minimizing temperature gradients.

We achieve temperature stability to within 1 mK on the timescale of weeks. However, ambient changes in temperature outside the box can cause the temperature control system to introduce changing temperature gradients inside the box, which are the present limiting factor for drift.

\begin{figure}
\includegraphics[width=176mm]{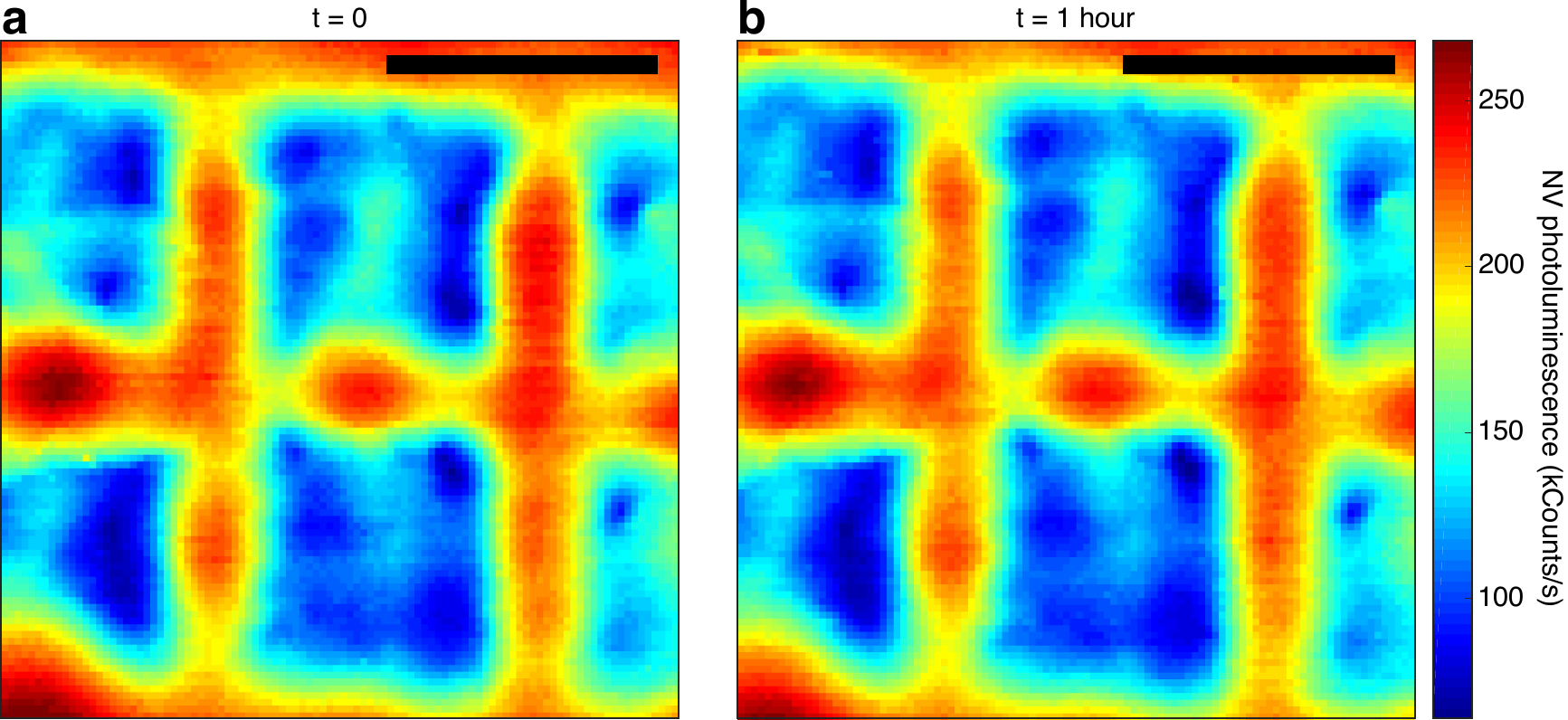}
\captionsetup{justification=raggedright, singlelinecheck=false}
\caption{NV-photoluminescence (PL)-based image registration for NV-sample drift correction. \textbf{a} and \textbf{b} are subsequent PL scans, taken 1 hour apart, of the nanopatterned sample in Fig.~\ref{CondImaging} of the main text. An image registration algorithm calculates an $(x,y)$ shift of $(-0.5, 4.5)$ nm. Scale bars are 400 nm. The spatial PL modulation results from several factors that depend on the NV's position relative to the tip: the Purcell effect, a change in the NV's dielectric environment, and an interferometric effect between the incoming excitation laser and the tip-reflected laser  \cite{Israelsen2014}. The large PL modulation is consistent with small ($\sim$10-20 nm) NV-sample separation.}
\label{PLRegimage}
\end{figure}

In order to correct for these drifts, we employ image registration every 1-2 hours, using either a topographic AFM image or an NV PL image (the one with the sharper features is chosen for image registration). 
Two example PL images taken an hour apart in time are shown in Fig.~\ref{PLRegimage}. We use an image registration algorithm that allows for subpixel correction to $\sim$ 1 nm precision \cite{Guizar-Sicairos2008,Pelliccione2014}. 

\section*{SUPPLEMENTARY NOTE 3: OPTIMAL $T_1$ MEASUREMENT}

For the exponentially decaying signal $S_{\text{diff}}=C \exp(-t/T_1)$, the most time-efficient method of measuring $T_1$ is by acquiring data at $\tau = 0$ and $\tau \approx 0.7 ~T_1$, which we show here.

Consider a weighted, least-squares linear fit for data with relationship $y_i = A + B x_i$ with weights $w_i = 1 / \sigma_{y_i}^2$, where $\sigma_{y_i}$ is the standard deviation of the measured $y_i$ values. One can calculate, then, that the variance on the fitted value of B is

\begin{flalign}
\sigma_B^2 = \frac{\sum w_i}{\sum w_i \sum w_i x_i^2 - \left(\sum w_i x_i \right)^2}
\end{flalign}

If we linearize our signal $S_{\text{diff}}$, referred to now as $S$ for brevity, we find that

\begin{gather}
\ln(S_i) = \ln(C) - \Gamma \tau_i
\\
\sigma_{\ln(S_i)} = \frac{\sigma_S}{S_i}
\\
\sigma_{\Gamma}^2 = \frac{\sigma_S^2 \sum S_i^2}{\sum S_i^2 \sum S_i^2 \tau_i^2 - \left(\sum S_i^2 \tau_i\right)^2}
\end{gather}

We consider $\sigma_S^2 = \sigma_1^2 / N$, where $N$ is the number of repetitions and $\sigma_1^2$ is the per-shot measurement error, which we assert is the same for each $\tau_i$. Consider a per-shot overhead time $t_{\text{extra}}$ and dark times $\tau_i$ such that $N \sum(\tau_i + t_{\text{extra}}) = T$, with $T$ the total measurement time. Thus, one finds that

\begin{flalign}
\sigma_{\Gamma}^2 &= \frac{\sigma_1^2}{T} \frac{\sum e^{-2\Gamma\tau_i} \sum(\tau_i + t_{\text{extra}})}{\sum e^{-2\Gamma\tau_i} \sum  \tau_i^2 e^{-2\Gamma\tau_i} - \left(\sum \tau_i e^{-2\Gamma\tau_i}\right)^2}
\end{flalign}

\begin {figure}
 \includegraphics[width=132mm]{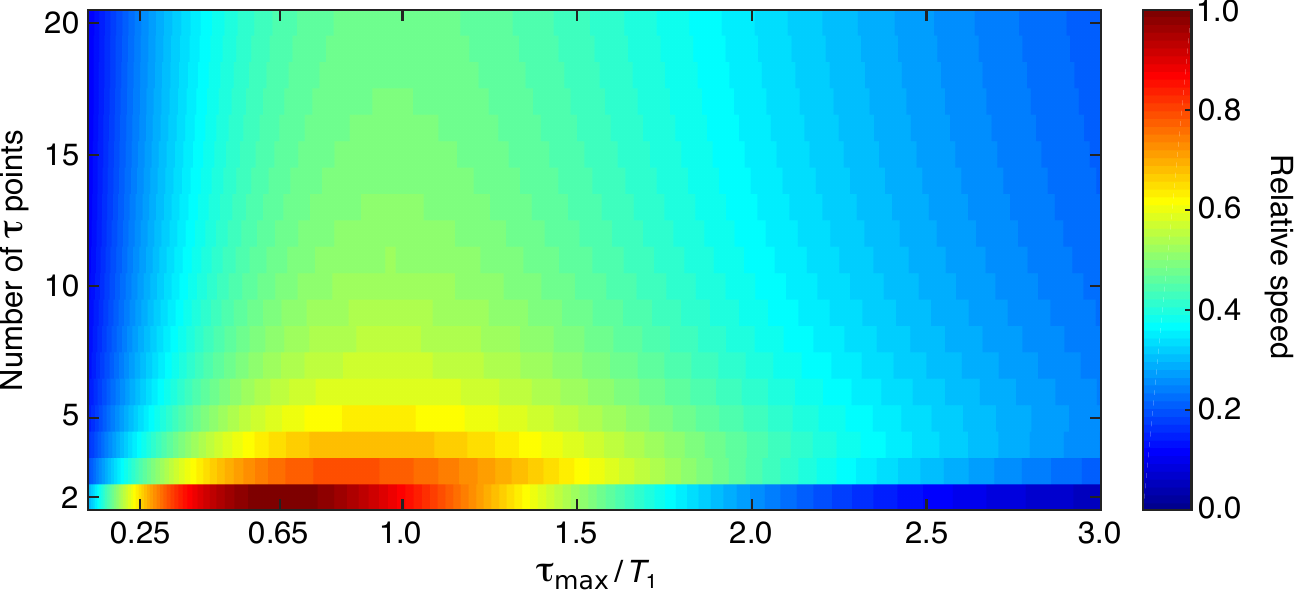}
\captionsetup{justification=raggedright, singlelinecheck=false}
\caption{$T_1$ measurement speed with different numbers of $\tau$ points spaced from $\tau=0$ to $\tau = \tau_{\text{max}}$. The optimal measurement sequence is a measurement at $\tau = 0$ and at $\tau = 0.65 ~T_1$. We assume a per-shot overhead time of 0.01 $T_1$. Deviating from $\tau \approx 0.7 ~T_1$ greatly increases measurement time, stressing the necessity of adaptive-$\tau$ measurements for imaging.}
\label{TauOptImage}
\end{figure}

In Fig.~\ref{TauOptImage} we plot the relative speed $(1 / \sigma_{\Gamma}^2)$ for $N$ linearly spaced $\tau$ values between $\tau = 0$ to $\tau_{max}$, as a function of $N$ and $\tau_{max}$. The plot shows a clear maximum in measurement speed at $\tau \approx 0.7~ T_1$. Fig.~\ref{TauOptImage} thus emphasizes the importance of an adaptive-$\tau$ algorithm when imaging over an area where $T_1$ varies. For example, in the area imaged in Fig.~\ref{CondImaging} of the main text $T_1$ varies from 0.5 to 5 ms.

\section*{SUPPLEMENTARY NOTE 4: SPIN-TO-CHARGE READOUT ON A SHALLOW NV CENTER}

\begin{figure}
\includegraphics[width=138mm]{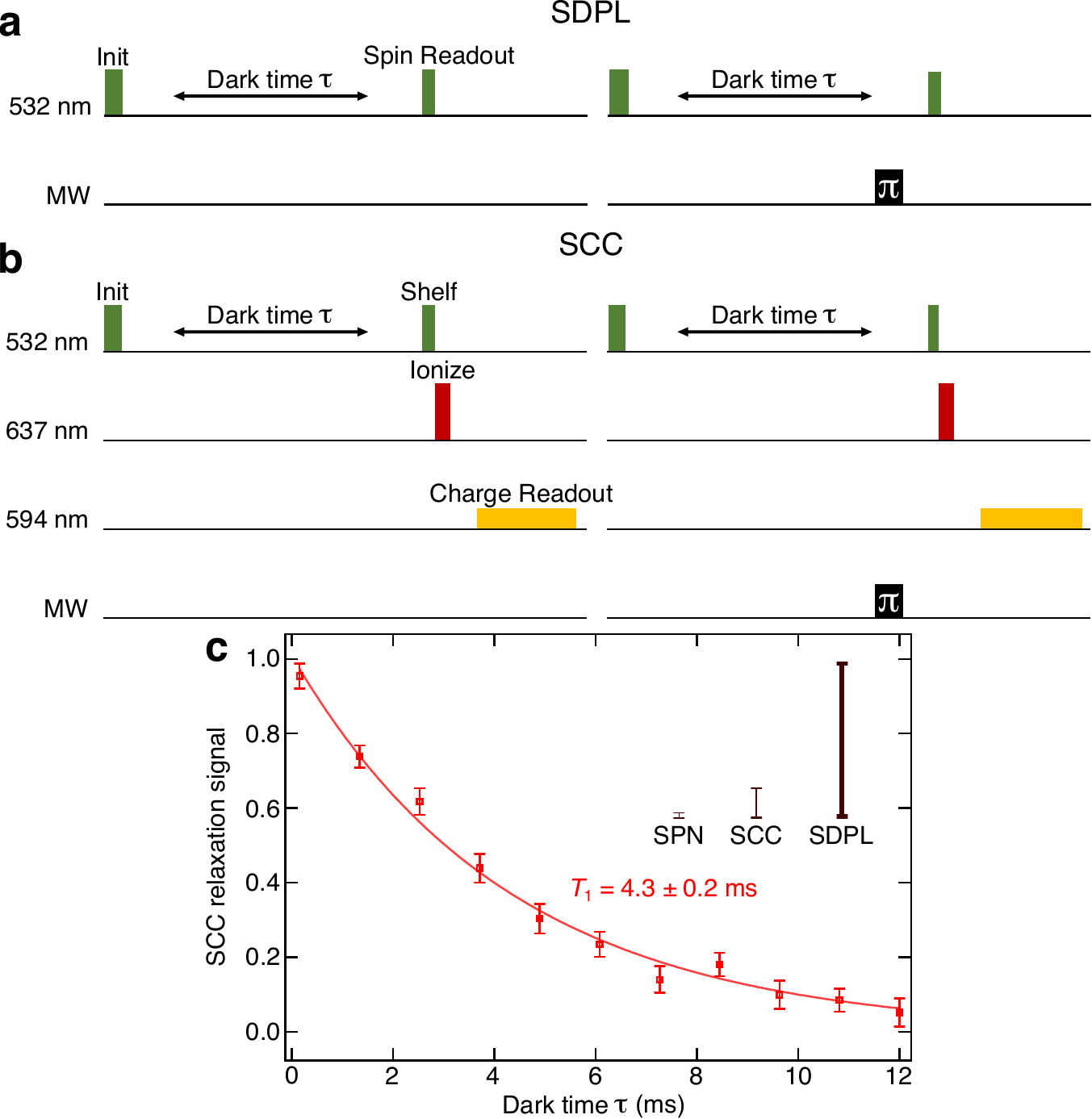}
\captionsetup{justification=raggedright, singlelinecheck=false}
\caption{Spin-to-charge conversion (SCC) $T_1$ measurement on a shallow NV center. \textbf{a} Conventional, spin-dependent photoluminescence (SDPL) $T_1$ measurement scheme. The $\pi$ pulse swaps the $|{0}\rangle$ and $|{-1}\rangle$ populations immediately before readout, as explained in Eq.~\ref{Seqn}. \textbf{b} Same preparation sequence as \textbf{a} but readout uses SCC. To convert spin to charge, we do a $\sim 300\text{-}\mu$W, 60-ns green pulse to shelve $|{\pm1}\rangle$ into the singlet, immediately followed by a $\sim 30$-mW, 40-ns red pulse in order to discriminately ionize $|{0}\rangle$ without touching the singlet population. We then perform charge-state readout with a $\sim 3\text{-}\mu$W, 500-$\mu$s yellow pulse which discriminately excites NV$^{-}$. \textbf{c} SCC $T_1$ measurement on a shallow NV center. The measurement took 30 minutes. Plotted error bars represent the measured standard error, and are 5x larger than the spin projection noise (SPN) limit. The inset shows the relative magnitude of typical SPN, SCC, and SDPL error (1, 5, 25). Nominally, to achieve the same error the SCC measurement is 25x faster than SDPL.}
\label{SCCPlot}
\end{figure}

For Fig.~\ref{CondImaging}c of the main text we implement spin-to-charge conversion (SCC) readout \cite{Shields:2015ik} on a shallow NV center with a measurement sensitivity 5x the spin projection noise (SPN) limit. Fig.~\ref{SCCPlot}b and Fig.~\ref{SCCPlot}c illustrate the SCC measurement scheme and show an example SCC $T_1$ measurement that is 5x the SPN limit. 

The SPN limit is the result of a Bernoulli distribution: each projective measurement will result in a success $|m_s = 0\rangle$ or a failure $|m_s = \pm 1\rangle$. We make a measurement at $\tau \approx 0.7 ~ T_1$, for which the spin population is roughly split in half between $|m_s=0\rangle$  and $|m_s=\pm1\rangle$. If the population is split in half between the two outcomes, this results in a standard deviation of 1/2. Since each measurement of $S_{\text{diff}}$ is the difference of two measurements ($S$ and $S_{\text{swap}}$), whose errors add in quadrature, the SPN-limited standard error $\sigma_{\text{SPN}}$ of $N$ measurements of $S_{\text{diff}}$ is 

\begin{equation}
\sigma_{\text{SPN}} = \frac{1}{\sqrt{2N}}
\end{equation}

More carefully accounting for the 3-level system yields a 5\% smaller SPN limit. 

In Fig.~\ref{CondImaging}c of the main text and in Fig.~\ref{SCCPlot} we experimentally measure $\sigma_{\text{SCC}} \approx 5 / \sqrt{2N} = 5~ \sigma_{\text{SPN}}$. For spin-dependent photoluminescence (SDPL) readout, we typically measure $\sigma_{\text{SDPL}} = 25 ~\sigma_{\text{SPN}} = 5 ~\sigma_{\text{SCC}}$, which is consistent with photon shot noise for the experimental parameters of a 400-ns long readout, $|0\rangle$-state PL of 180 kCounts/s, and PL contrast of 30\%  between $|0\rangle$ and $|\pm1\rangle$ states. Further, note that the SPN value quoted above assumes perfect initial polarization into NV$^{-}$ and $|m_s = 0\rangle$; in practice imperfect spin and charge polarization increases our experimental error by a factor of $\sim 1.3$. Thus, in reality our measured readout error $\sigma_{\text{SCC}} \approx 4 ~\sigma_{\text{SPN}}$.

\end{document}